\documentclass[twocolumn,showpacs,preprintnumbers,amsmath,amssymb]{revtex4}

\usepackage{graphicx}

\begin{document}

\title{Determination of the Intershell Conductance in Multiwalled Carbon Nanotubes}

\author{B. Bourlon,$^1$ C. Miko,$^2$ L. Forr\'{o},$^2$ D.C. Glattli,$^{1,3}$ A. Bachtold$^{1*}$}

\address{
$^1$ Laboratoire Pierre Aigrain, Ecole Normale Sup\'{e}rieure, 24
rue Lhomond, 75231 Paris 05, France. $^2$ EPFL, CH-1015, Lausanne,
Switzerland. $^3$ SPEC, CEA Saclay, F-91191 Gif-sur-Yvette,
France.}

\begin{abstract}

We report on the intershell electron transport in multiwalled
carbon nanotubes (MWNT). To do this, local and nonlocal four-point
measurements are used to study the current path through the
different shells of a MWNT. For short electrode separations
$\lesssim$ 1 $\mu$m the current mainly flows through the two outer
shells, described by a resistive transmission line with an
intershell conductance per length of $\sim$(10
k$\Omega$)$^{-1}$/$\mu$m. The intershell transport is tunnel-type
and the transmission is consistent with the estimate based on the
overlap between $\pi$-orbitals of neighboring shells.

\end{abstract}

\vspace{.3cm} 

\date{ \today}
\maketitle

The specific geometry of MWNTs that consist of several nested
cylindrical graphene shells (Fig. 1(a)) has motivated intense work
to understand their mechanic and electronic properties
\cite{Schonenberger}. MWNTs are found to conduct charges
exceptionally well with a conductivity better \cite{Frank} or
comparable \cite{Bachtold} to that of low-resistivity metals. This
high conductivity is attributed to the low level of diffusion
experienced by conducting modes along the shells. These modes are
characterized by long wavelengths in the circumference direction
(Fig. 1(b)) as predicted from the electronic structure
calculation, which is based on the delocalisation of $\pi$-orbital
states along the shell surface (Fig. 1(c)).

In MWNTs, neighboring shells are most often incommensurate. The
resulting lack of periodicity is expected to inhibit the charge
delocalization in the radial direction. The intershell conduction
is thus governed by hopping and depends on the  $\pi$-orbital
overlap of nearby shells (Fig. 1(c)). Therefore, the intershell
conduction mechanism is thought to be significantly different
compared to that in the sheet plane. However, the intershell
resistance remains to be experimentally quantified.

\begin{figure}
\includegraphics{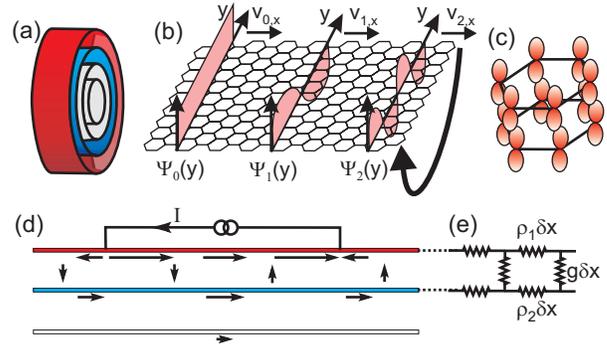}
 \caption{
(color online). Description of multiwalled nanotube. (a) A MWNT
which consists of 4 shells. (b) A graphene sheet which gives a
MWNT shell when rolled-up. Also shown is the slowly varying
transverse part $\propto exp(ikr)$ of the electronic wave
function. $\Psi_0$ corresponds to the two crossing subbands at the
$K$ point and $\Psi_1$ and $\Psi_2$ to the next successive
subbands. (c) A hexagon of carbon $\pi$-orbitals. The two hexagons
belong to two nearby incommensurate shells. The intershell
resistance depends on the orbital overlap. (d) Current pathways.
The 3 outermost shells are represented by lines of different
colors. When current is externally applied, the current enters and
exits the MWNT through the outermost shell. Importantly, part of
$I$ flows through to the second shell, so that the current pathway
penetrates beside the region lying between the bias electrodes.
(e) Resistive transmission line, which models the two outermost
shells.
 }
\end{figure}

As a first indication for the intershell resistance one could
think of relying on the interlayer resistivity of pure graphite
$\sim10^{-5}$ $\Omega$m. But, this value is ultra sensitive to the
material quality and the interlayer separation, so that the
resistivity is easily increased up to $\sim10^{-3}$ $\Omega$m
\cite{Dutta,Primak,Uher,Matsubara}. On the theoretical side, many
groups have studied the intershell conduction but results differ
dramatically depending on the model hypotheses
\cite{Sanvito,Maarouf,Roche,Yoon,Kim,Ahn,Hansson,Uryu,Triozon}.
For example, zero intershell transport is predicted for an
infinitely long and perfect tube, since both the energy and the
Bloch wave vector have to be conserved \cite{Yoon}. In contrast,
it has been shown that the intershell transmission is large when
electrons are injected as localized wave packets from outside the
shell \cite{Roche}. The attenuation length, which is the length
necessary for a charge flowing along a shell to propagate into the
next shell, was found to be very short $L_a \sim$1-10 nm
\cite{Roche}. Experimentally, some results indicating current
anisotropy have been reported. It has been suggested that the
current flows through one or a few of the outermost shells
\cite{Frank,Bachtold} when the current is injected in the outer
shell.

We report here a new method to access the electronic paths in
MWNTs which enables the estimation of the linear intershell
resistance. Using four-point measurement techniques, the voltage
drop is measured between electrodes located inside or outside the
region lying between the current biased electrodes. Surprisingly,
a significant nonlocal voltage drop is observed. Such a result has
been reported previously in a proceeding \cite{Collins3} and
attributed to multiple shells conduction with the intershell
electron pathways occurring near defects or tube end caps. We show
here that the nonlocal voltage drop decreases exponentially with
distance. Moreover, the local voltage measured in a standard
four-probe configuration drops when the distance between the
current electrodes is increased. These results are in agreement
with a model which considers conduction through the two outermost
shells and treats them as a resistive transmission line (Fig.
1(d,e)). In such a model, the intrashell resistance is $\sim$ 10
k$\Omega$/$\mu$m and the intershell conductance is $\sim$(10
k$\Omega$)$^{-1}$/$\mu$m. This latter value is in agreement with
the estimate based on electrons tunnelling through atomic orbitals
of nearby shells while taking into account conservation of energy
but not momentum. From this value, and the 0.34 nm intershell
separation, we deduce a radial resistivity for MWNTs of $\sim$1
$\Omega$m, much larger than that of bulk graphite.

\begin{figure}
\includegraphics{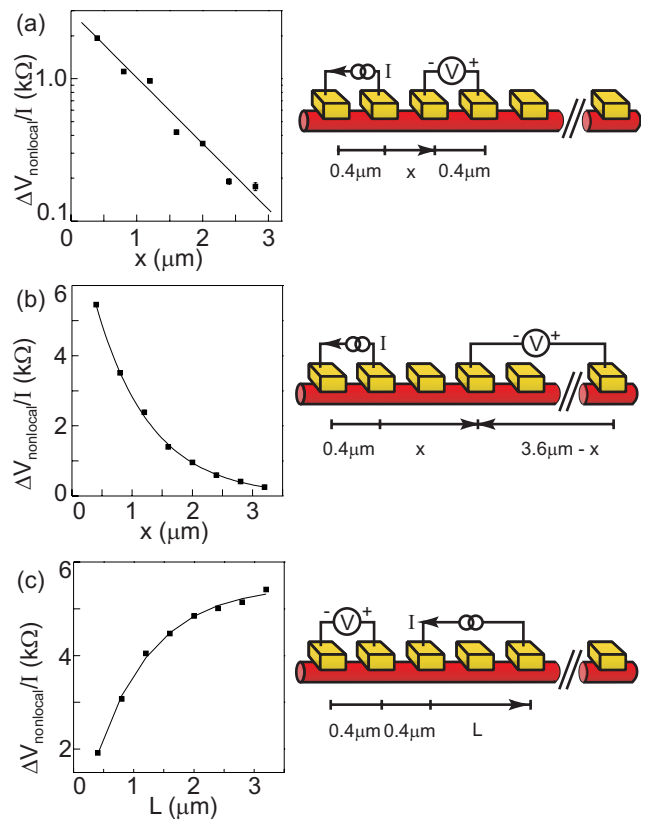}
 \caption{
(color online). Nonlocal voltage measurements on a MWNT
electrically addressed by 11 electrodes. The schematics show the 5
$\mu$m long MWNT. The diameter is 17 nm, which corresponds to
about 20 shells. The MWNT, synthesized by arc-discharge
evaporation and carefully purified \cite{Bonard}, was dispersed
onto a 500 nm oxidized Si wafer from a dispersion in
dichloroethane. Cr/Au electrodes were patterned above the tube by
electron beam lithography. (a) $\Delta V_{nonlocal}$ as a function
of the separation between the current and the voltage electrodes.
$\Delta V_{nonlocal}/I$ is plotted on a logarithmic scale. The
straight line corresponds to a decay length of 0.94 $\mu$m. Data
are taken at 250 K. $\Delta V_{nonlocal}/I$ is measured in the
linear regime with $eV=eIR_{2P}$ below $kT$, $R_{2P}$ being the
two-point resistance. (b) Nonlocal voltage as a function of the
tube length. The voltage reference is taken at the extremity of
the electrode series. The continuous curve is an exponential decay
fit with a decay length of 0.92 $\mu$m. (c) $\Delta V_{nonlocal}$
as a function of the separation between the current electrodes.
The continuous curve is proportional to $1-exp(-L /L_a )$ with
$L_a=$ 0.94 $\mu$m.
 }
\end{figure}

We look first at the measurements of the nonlocal voltage $\Delta
V_{nonlocal}$ (schematic of Fig. 2(a)). $\Delta V_{nonlocal}$ is
linear with the applied current $I$ at low bias. Such a finite
nonlocal voltage at room temperature is a signature of systems
possessing complex current pathways and is particularly evident in
anisotropic materials such as high-temperature superconductors
\cite{Busch}. An example of current pathway is illustrated in Fig.
1(d). Interestingly, the values of the nonlocal resistance $\Delta
V_{nonlocal}/I$ are comparable for all $\sim$ 40 studied MWNTs,
which suggests similar current pathways for different MWNTs.
Indeed, we get $V_{nonlocal}/I=430 \pm$90 $\Omega$  for devices
with 400 nm electrode separations.

Now, we turn to the spatial dependence of the nonlocal voltage.
For this purpose, 11 electrodes were attached on a single MWNT
with about 20 shells. This allowed us to probe the  $\Delta
V_{nonlocal}$ dependence on the lengths $x$, $L$ and $d$, where
$x$ is the separation between the voltage and current electrodes,
$L$ the separation between the current electrodes and $d$ the
separation between the voltage electrodes. Figure 2(a) shows that
$\Delta V_{nonlocal}/I$ decreases exponentially with $x$ with a
decay length of 0.94 $\mu$m. We note that the current is zero far
from the current electrodes such as at the tube end caps, showing
that those play a negligible role in the current pathway. To
verify this measurement, the voltage profile is measured along the
tube with the farthest electrode taken as voltage reference (Fig.
2(b)). Again, the voltage decreases exponentially with a decay
length equal to 0.92 $\mu$m.

When the separation $L$ between the current biased electrodes is
increased,  $\Delta V_{nonlocal}$ gets larger as shown in Fig.
2(c). The spatial dependence is an exponential with a
characteristic length of 0.94 $\mu$m. This measurement shows that
more current is injected in inner shells for longer $L$. Indeed,
only the current in the inner shells leads to current though the
outermost shell in the nonlocal region (see Fig. 1(d)).

\begin{figure}
\includegraphics{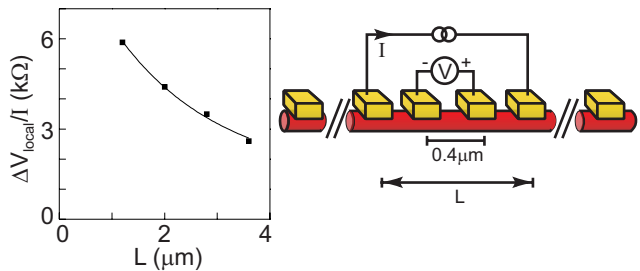}
 \caption{
(color online). Local voltage as a function of separation between
the current biased electrodes. $L$ is symmetrically increased with
respect to the center between the voltage electrodes. The
continuous curve is exponential with a decay length equal to 2
$\mu$m.
 }
\end{figure}

We now turn our attention to the traditional four-point
configuration with local voltage  $\Delta V_{local}$ measurements.
Fig. 3 shows that  $\Delta V_{local}$ is not independent on $L$
the distance between the current probes as would be expected for a
standard four-point measurement, but instead  $\Delta V_{local}$
increases as $L$ is reduced. The spatial dependence is consistent
with an exponential decay with a characteristic length of 2
$\mu$m, which is twice the value found above. This result points
to the same finding as in the last paragraph, namely that the
current injected into the inner shells increases with the
separation between the current biased electrodes.

The current injection into the inner shells might be related to
the electrodes. The MWNT structure could be modified by the
process used for the fabrication of electrodes on top of the tube.
However, measurements using a different device configuration,
where the MWNT was deposited on top of existing electrodes, give
similar results. This suggests that the fabrication process does
not result in any MWNT damage that modifies its transport
properties. An alternative effect related to electrodes might be
that their presence is at the origin of the length dependence of
$\Delta V_{local}$ and  $\Delta V_{nonlocal}$. Indeed, the decay
length is comparable to the electrode separation. To control for
this effect, devices were fabricated with 6 electrodes separated
by 400 nm except between the fourth and the fifth electrodes that
were separated by 1100 nm. The two middle electrodes are current
biased. The $\Delta V_{nonlocal}$ measured at $x=$400 nm is lower
than that measured at $x=$1100 nm with a ratio consistent with
measurements on MWNTs contacted by electrodes separated always by
400 nm. Therefore, the observed exponential dependences are an
intrinsic property of the MWNTs we study. In addition, the metal
electrodes do not provide additional pathways for the current to
flow, short-circuiting the outer shell. Indeed, $\Delta
V_{nonlocal}/I$ is observed to not depend on a low or large
transmission at the contacts \cite{rem2}.

We discuss now the number of current carrying shells. The nonlocal
voltage on the outermost shell is seen to decay exponentially on
the scale of a micron. Thus, a fraction of the current leaves the
outermost shell to penetrate into the inner shells over a typical
length scale of $L_a \sim$ 1 $\mu$m. The same phenomenon is
expected to occur between the second shell and deeper shells. Due
to the exponential dependence, only a small fraction of the
current is thus expected to reach the third and the deeper shells
when the separation between the current bias electrodes is lower
than $\sim 1 \mu$m.

Now, we consider a model based on a resistive transmission line
for a quantitative analysis of the results (Fig. 1(e)). On a
$\delta x$ length, the intershell conductance is $g\delta x$ and
the intrashell resistances are $\rho _1\delta x$ and  $\rho
_2\delta x$. For an infinite long MWNT, it gives

\begin{eqnarray}
\frac{\Delta V_{nonlocal}}{I}  &=&  \frac{g\rho^2_1
L^3_a}{2}\exp(\frac{-x}{L_a})\\ \nonumber & \cdot&
\left(1-\exp(\frac{-d}{L_a})\right)\left(1-\exp(\frac{-L}{L_a})\right)
\end{eqnarray}
\begin{equation}
\frac{\Delta V_{local}}{I}=g\rho_1 L^3_a
\left(\frac{\rho_2d}{L_a}+2\rho_1
\textrm{sh}(\frac{d}{2L_a})\exp(\frac{-L}{2L_a})\right)
\end{equation}

with $L^{-1}_a=\sqrt{g(\rho_1+\rho_2)}$. Equation (1) predicts an
exponential form of $\Delta V_{nonlocal}/I$ as a function of $x$
and $L$ with the same characteristic length $L_a$, in excellent
agreement with experiments. $\Delta V_{local}/I$ is expected in
equation (2) to depend exponentially on $L$, this time with a
characteristic length $2L_a$, which agrees again with experiments.

We estimate here $\rho _1$, $\rho _2$ and $g$ by taking $L_a=$0.93
$\mu$m and comparing equations (1,2) to the measured values of
$\Delta V_{nonlocal}/I$ for $x=L=d=$0.4 $\mu$m and $\Delta
V_{local}/I$ for $d=L/3=$0.4 $\mu$m.  However, $\Delta
V_{nonlocal}/I$ or $\Delta V_{local}/I$ are not constant for
different sections of the tube. These variations are attributed to
local inhomogeneities in the electronic diffusion along the shell
and/or to the imprecision in $x$, $L$ and $d$ due to the finite
width 200 nm of the electrodes. Indeed, the electron transmission
is probably non uniform along the tube/electrode interface and
occurs at one or a few more preferential places. For a more
accurate determination, we average over the 8 independent
measurements obtained by translating the set of active probes by
one unit each time. We get $\Delta V_{nonlocal}/I=$780$\pm$170
$\Omega$ and $\Delta V_{local}/I=$4.8$\pm$0.4 k$\Omega$. This
gives $\rho _1 \sim$22 k$\Omega$/$\mu$m,   $\rho _2 \sim$1
k$\Omega$/$\mu$m and $g\sim$(20 k$\Omega$)$^{-1}$/$\mu$m. Taking
these values and assuming the resistivity of the third shell $\rho
_3 =\rho _2$, we estimate the transport contribution of the third
shell to be less than 10$\%$ for a length shorter than 5 $\mu$m.
This supports the double shell conduction analysis used here.

The same analysis is made for 8 different MWNTs which are 6-23 nm
thick and which are connected by at least 7 electrodes. We obtain
$L_a=$0.4-1 $\mu$m, $\rho _1=$6-25 k$\Omega$/$\mu$m, $\rho
_2=$0.05-2$\rho _1$ and $g=$ (3.7-20 k$\Omega$)$^{-1}$/$\mu$m. We
note that $\rho _1$ is consistent with the MWNT resistance per
length previously obtained \cite{Bachtold2}. In addition, the
estimate for $\rho _2$ is very sensitive to variations of $\Delta
V_{nonlocal}/I$ and $\Delta V_{local}/I$. However, $\rho _2$ is
most often found lower than $\rho _1$, which suggests stronger
diffusion in the outermost shell than in deeper shells. A possible
explanation is that inner shells are protected by the outermost
shell, which may be in direct contact with some adsorbed molecules
or which may have been degraded during manipulation.

\begin{figure}
\includegraphics{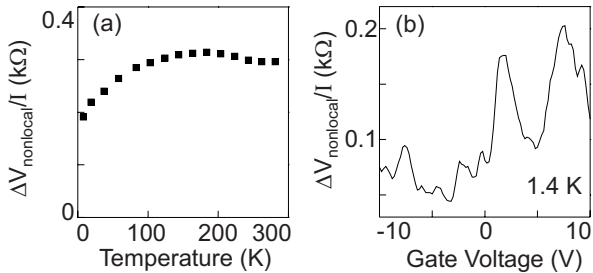}
 \caption{
Nonlocal voltage at low temperature. (a)  $\Delta V_{nonlocal}$ as
a function of temperature measured from 8 to 280 K. The MWNT
diameter is 7 nm and the electrode separations are 450 nm. $\Delta
V_{nonlocal}$  is averaged over the gate voltage. (b) $\Delta
V_{nonlocal}$ as a function of gate voltage.
 }
\end{figure}

The intershell conductivity $g\sim$(10 k$\Omega$)$^{-1}$/$\mu$m is
relatively large, when compared to most theoretical expectations
\cite{Sanvito,Maarouf,Roche,Yoon,Kim,Ahn,Hansson,Uryu,Triozon}. As
for the dependence on temperature $T$, Fig. 4(a) shows a weak
variation $\Delta V_{nonlocal}/I$ versus $T$ above 10 K, which
suggests that the intershell conduction is not thermally
activated. At lower temperatures, quantum effects appear as seen
in the sensitivity of $\Delta V_{nonlocal}/I$ to a gate voltage
which may be attributed to electronic interference and/or Coulomb
interaction \cite{Buitelaar} (Fig. 4(b)). Importantly, the
nonlocal voltage is well above zero showing that the intershell
transmission is far from being blocked for $T$ above 1.4 K.

The intershell conductance $g$ is compared here to the predictions
of tunnelling conductance $G_{atom}$ through the $\pi$-orbital
overlap between 2 atoms of nearby shells. Assuming elastic
tunneling, $G_{atom}$ is given by \cite{Chen}


\begin{equation}
G_{atom}=\frac{4\pi e^2}{\hbar}N^2_{atom}E^2_{bin}
\end{equation}

with $E_{bin}$ the binding energy due to electronic delocalisation
and $N_{atom}=\frac{2n}{hv_f}\frac{S}{2\pi r}$ the density of
state per atom. $S=$ 2.6 $\textrm{\AA}^2$ is the surface occupied
by one atom, $r$ the shell radius, $v_f=$ 10$^6$ ms$^{-1}$ the
Fermi velocity and $n=$ 10-20 the number of modes due to doping
\cite{Kruger}. $G_{atom}$ is related to $g$ through
$g=G_{atom}\frac{2\pi r}{S}$. Using $g=$ (10
k$\Omega$)$^{-1}$/$\mu$m, $n=$ 15 and $r$=5 nm, we obtain that
$E_{bin}$ is on the order of 25 meV, in agreement with reported
values for graphite. Indeed, the binding energy due to electronic
delocalisation has been reported to contribute significantly to
the interplanar cohesive energy, which has been theoretically
estimated \cite{Schabel,Charlier} at $\sim 25$ meV and measured
\cite{Benedict} around $\sim 35$ meV. We note that Eq. (3) gives a
rather rough estimate for the intershell conduction. Indeed,
$G_{atom}$ is expected to be reduced due to shell
incommensurability or Bloch wave vector conservation which is
obeyed only partially because of disorder, finite tube length and
tube deformation due the substrate roughness. On the other hand,
the tube deformation can locally enhance the orbital overlap and
thus $g$. However, a quantitative estimate of these effects is
difficult at this stage.

The measure of the intershell conductance represents an important
step in the characterisation of basic MWNT properties. We have
found that the intershell conductance is largely independent of
temperature and is consistent with tunnelling through orbitals of
nearby shells. We have also shown that the current flows mainly
along individual shells that are rather efficiently insulated from
each other in spite of the short 0.3 nm shell separation. The
control of charge pathways on the angstrom scale is obviously
promising for future experiments or applications. For example, the
effect on transport of the Coulomb interaction between
neighbouring shells can be studied. It is also possible to
fabricate an intramolecular field-effect transistor based on
engineered MWNTs \cite{Collins1,Bourlon} using a thin-diameter
semiconducting shell that is gated by the next shell.

We thank B. Placais and C. Baroud for discussions and C. Delalande
for support. LPA is CNRS-UMR8551 associated to Paris 6 and 7. The
research has been supported by the DGA, ACN, sesame, the Swiss
National Science Foundation and its NCCR "Nanoscale Science".

$^{*}$ corresponding author: bachtold@lpa.ens.fr


\begin{references}

\bibitem{Schonenberger}
C. Schonenberger, L. Forro, Phys. World \textbf{13}, 37 (2000).


\bibitem{Frank}
S. Frank $et$ $al.$, Science {\bf 280}, 1744 (1998).

\bibitem{Bachtold}
A. Bachtold $et$ $al.$, Nature {\bf 397}, 673 (1999).

\bibitem{Dutta}
A.K. Dutta, Phys. Rev. \textbf{90}, 187 (1953).

\bibitem{Primak}
W. Primak, Phys. Rev. \textbf{103}, 544 (1956).

\bibitem{Uher}
C. Uher, L.M. Sander, Phys. Rev. B 27, 1326 (1983).

\bibitem{Matsubara}
K. Matsubara, K. Sugihara, T. Tsuzuku, Phys. Rev. B \textbf{41},
969 (1990).

\bibitem{Sanvito}
S. Sanvito \textit{et al.}, Phys. Rev. Lett. \textbf{84}, 1974,
(2000).

\bibitem{Maarouf}
A.A. Maarouf, C.L. Kane, E.J. Mele, Phys. Rev. B \textbf{61},
11156 (2000).

\bibitem{Roche}
S. Roche \textit{et al.}, Phys. Rev. B \textbf{64}, R121401
(2001).

\bibitem{Yoon}
Y.G. Yoon, P. Delaney, S. G. Louie, Phys. Rev. B \textbf{66},
073407 (2002).

\bibitem{Kim}
D.H. Kim, K.J. Chang, Phys. Rev. B \textbf{66}, 155402 (2002).

\bibitem{Ahn}
K.H. Ahn \textit{et al.}, Phys. Rev. Lett. \textbf{90}, 026601
(2003).

\bibitem{Hansson}
A. Hansson, S. Stafstrom, Phys. Rev. B \textbf{67}, 075406 (2003).

\bibitem{Uryu}
S. Uryu, Phys. Rev. B \textbf{69}, 075402 (2004).

\bibitem{Triozon}
F. Triozon \textit{et al.}, Phys. Rev. B \textbf{69}, R121410
(2004).

\bibitem{Collins3}
P.G. Collins \textit{et al.}, \textit{Proc. 14th Int. Winter
School on Electronic Properties of Novel Materials} (Am. Inst.
Phys., New York, 2000).

\bibitem{Busch}
R. Busch et al., Phys. Rev. Lett. \textbf{69}, 522 (1992).

\bibitem{rem2}
The contact resistance is roughly estimated by subtracting the
four-point measurement from the two-point measurement, which
varies from 5 k$\Omega$ to 3 M$\Omega$ for the MWNT of Fig. 2,3.


\bibitem{Bachtold2}
A. Bachtold\textit{ et al.}, Phys. Rev. Lett. \textbf{84}, 6082
(2000).

\bibitem{Buitelaar}
M.R. Buitelaar \textit{et al.}, Phys. Rev. Lett. \textbf{88},
156801 (2002).

\bibitem{Chen}
C.J. Chen, \textit{Introduction to Scanning Tunnelling Microscopy}
(Oxford University Press, 1993).

\bibitem{Kruger}
M. Kruger \textit{et al.}, Appl. Phys. Lett. \textbf{78}, 1291
(2001)

\bibitem{Schabel}
M.C. Schabel, J.L. Martins, Phys. Rev. B \textbf{46}, 7185 (1992).

\bibitem{Charlier}
J.C. Charlier, X. Gonze, J.P. Michenaud, Europhys. Lett.
\textbf{28}, 403 (1994).

\bibitem{Benedict}
L.X. Benedict \textit{et al.}, Chem. Phys. Lett. \textbf{286}, 490
(1998).

\bibitem{Collins1}
P. G. Collins, M. S. Arnold, Ph. Avouris, Science {\bf 292}, 706
(2001).

\bibitem{Bourlon}
B. Bourlon \textit{et al.}, Phys. Rev. Lett. \textbf{92}, 026804
(2004).

\bibitem{Bonard}
J.M. Bonard \textit{et al}., Adv. Mater. \textbf{9}, 827 (1997).

\end{references}
\end{document}